\documentstyle[psfig,times]{l-aa}

\begin{document}

\thesaurus{03          
		   (02.12.1;   
			02.19.2;   
			03.20.5;   
			11.01.2;   
			13.18.1)}  
\title{Anisotropic [OIII] emission in radio loud AGN
\thanks{Based on observations obtained 
at the European Southern Observatory, La Silla, Chile, and at the Isaac
Newton Group of Telescopes, La Palma,
Canary Islands, Spain}}
\author{S. di Serego Alighieri\inst{1}, A. Cimatti\inst{1},
R.A.E. Fosbury\inst{2,3} and R. Hes\inst{4,5} }
\offprints{S. di Serego Alighieri}

\institute{
            Osservatorio Astrofisico di Arcetri,
               Largo E.Fermi 5,
               I-50125 Firenze,
               Italy;
			   sdiserego@arcetri.astro.it, acimatti@arcetri.astro.it
	\and
            Space Telescope--European Coordinating Facility,
			 Karl-Schwarzschild Str.2,
			 D-85748 Garching bei M\"unchen,
			 Germany;
			 rfosbury@eso.org
	\and
			Affiliated to the Astrophysics Division of the 
			Space Science Dept., European Space Agency
	\and
			Dunsink Observatory,
			School of Cosmic Physics,
			Castleknock, Dublin 15,
			Ireland
    \and
	        Present address: Faculty of Systems Engineering,
			Group of Information and Communication Technology,
			Delft Univ. of Technology,
			PO Box 5015,
			2600 GA Delft,
			The Netherlands;
			rhes@sepa.tudelft.nl
           }

\date{ Received .... 1997; accepted ..... }

\maketitle

\markboth{S. di Serego Alighieri et al.: Anisotropic [OIII] emission in
radio loud AGN}{}

\begin{abstract}
We present the results of spectropolarimetry of a sample of 7 powerful
radio galaxies and 2 quasars with $0.07<z<0.35$ obtained to detect
possible anisotropies in the [OIII] line emission, which could explain
the higher [OIII] luminosity of quasars than that of radio galaxies
within the framework of the Unified Model of radio-loud AGN. We
detect polarized [OIII] in 4 radio galaxies, consistent with the
possibility that a considerable fraction ($\sim 20\%$) of the
observed [OIII] emission is scattered in a similar way to the hidden
nuclear continuum. However, small but detectable rotation between the
polarization direction of the line and of the continuum in two radio
galaxies shows that the geometry of the emitting regions can be
different.

\end{abstract}
\section {Introduction}
Jackson \& Browne (1990, hereafter JB90) have found that the [OIII]
5007\AA~line luminosity of a sample of powerful radio galaxies (RG) 
from the 3CRR catalogue with
$0.15<z<0.85$ is lower by factors of 5--10 than that of a corresponding
sample of quasars (RQ), matched in redshift and extended radio power.
Lawrence (1991) also finds that, at a given radio power quasars and
broad line radio galaxies have stronger [OIII] emission than narrow line
radio galaxies in a sample of FRII 3CRR sources with $z<0.5$. Similarly,
Seyfert 1 galaxies have higher [OIII] luminosity than Seyfert 2,
for a given radio luminosity (Lawrence 1987). However Keel et al. (1994)
found no significant difference in the [OIII] luminosity between type 1
and type 2 objects for a sample of IRAS Seyfert galaxies selected for
their warm far infrared colours.
This difference in [OIII] luminosity between RQ and RG has been regarded as a
failure of the Unified Model for radio--loud AGN, since it was thought
that this forbidden line would be emitted only from regions at a large
enough distance from the nucleus to be unobscured by the material
causing the anisotropy in the featureless nuclear continuum and in the
permitted broad lines, and that therefore its luminosity should be the
same  for RQ and RG, if they are parent populations, as discussed in
Antonucci (1993).
More recently, however, Hes et al. (1993, hereafter HBF93) have found that the
[OII] 3727\AA~line luminosity is the same for matched samples of RG and
RQ in a redshift range similar to that of JB90 ($0.2<z<0.8$), consistent with
the prediction of the Unified Model. 

\begin{figure}[htbp]
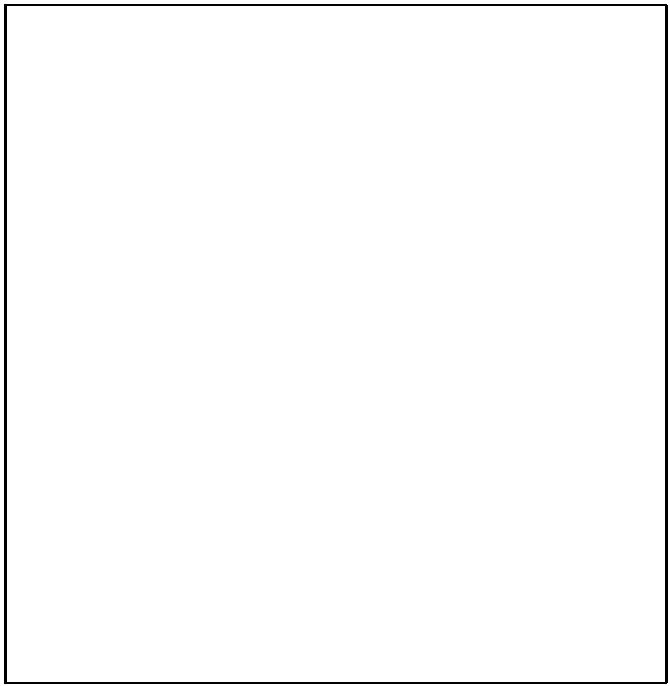

\picplace{9cm}
\caption{
Histograms of the [OIII] and [OII] line luminosities for radio galaxies
(thin line) and quasars (thick line) with 
$0.15<z<0.7$ in the complete sample of southern
2Jy radio sources of Tadhunter et al. (1993).
}
\label{fig1}
\end{figure}

These findings are confirmed by an analysis of the [OIII] and [OII] line
luminosities of the complete sample of 2Jy southern radio sources
observed by Tadhunter et al. (1993). Also in this case the average
[OIII] luminosity of the RQ with $0.15<z<0.7$ exceeds that of the
RG in the same redshift range by a factor of about 6, while the average
[OII] luminosities are about equal (see Fig.~\ref{fig1}).

HBF93 suggest that the partial
failure of the [OIII] test might be due to the fact that this line,
which has a higher critical density than [OII], might have a significant
component from the nuclear region, and thus be subject to pronounced
anisotropic obscuration. If this were the case, the [OIII]5007 line
would be partially polarized in RG, since its obscured component would
be scattered as we know happens for the broad permitted lines (Antonucci
1984, di Serego Alighieri et al. 1994). Therefore we have
performed a spectropolarimetric study of a sample of 7 RG to search for
polarized [OIII] line emission. We have also examined the polarization
of [OII] to see if it behaves differently from [OIII]. Finally
we have observed 2 RQ, in which there should be a much smaller scattered
fraction.

\begin{table*}
\caption{The sample and the observations}
\label{obs}
\[
\begin{tabular}{llllllll}
\hline
\noalign{\smallskip}
Object & Type & z & Radio P.A. & Opt. P.A. & Slit P.A. & Slit length & Exp. Time \\
& & & (deg.) & (deg.) & (deg.) & (arcsec) & (min.) \\
\noalign{\smallskip}
\hline
\noalign{\smallskip}
3C 227 & BLRG & 0.0861 & 84 & 119 & 90 & 15.2 & 4$\times$15 + 4$\times$15 \\
3C 327 & RG & 0.1039 & 100 & 98 & 127 & 9.1 & 4$\times$10 \\
0806$-$10 & RG & 0.110 & 19 & 40 & 30 & 7.5 & 4$\times$20 + 4$\times$20 \\
1549$-$79 & RG & 0.150 & 68 & 169 & 169 & 14.0 & 4$\times$20 + 4$\times$20 \\
3C 273 & RQ & 0.158 & 42 & 42 & 42 & 14.0 & 4$\times$3 + 4$\times$3 \\
3C 196.1 & RG & 0.198 & 42 & 52 & 54 & 13.4 & 4$\times$20 + 4$\times$20 \\
3C 180 & RG & 0.220 & 171 & 28 & 34 & 14.0 & 4$\times$15 + 4$\times$12 \\
1151$-$34 & BLRG & 0.258 & 70 & round & 90 & 12.8 & 4$\times$15 \\
1355$-$41 & RQ & 0.313 & 124 & unresolved & 90 & 12.8 & 4$\times$10 + 4$\times$10 \\
\noalign{\smallskip}
\hline
\end{tabular}
\]
\end{table*}

\section{Observations and analysis} 

The sample of RG and RQ was selected with the following criteria:
1. the redshift should be between 0.07 and 0.35, so that the emission
lines of [OIII] at 4959 and 5007\AA~and of [OII] at 3727\AA~ fall in the
range where EFOSC1, the instrument used, has good efficiency;
2. the coordinates should be in the range $6^h\leq R.A.\leq 16^h$ and
$Dec.\leq 10^{\circ}$, to be observable from La Silla in the allocated
period;
3. the radio source should be in the 3CR (Spinrad et al. 1985) or in the
2Jy catalogue (Wall \& Peacock 1985), which are the most completely
identified catalogues of powerful radio sources covering most of the sky. 

We have observed 9 of the 13 radio sources, which fulfill the above
criteria. Of the remaining 4 sources, 3C~287.1 (an N galaxy at z=0.2159)
has been observed spectropolarimetrically by Antonucci (1984), and the
others (3C~198, a RG at z=0.0815, 0736$+$01, a RQ at z=0.191, and
0859$-$25, a RG at z=0.305) were not observed for lack of time. However
the 9 objects observed were not selected out of the sample of 13
for any specific reason. Therefore our observations should
give an unbiased view of the problem of [OIII] anisotropy in radio
loud AGN.

The object 1151$-$34 was classified as a quasar by Wall \& Peacock
(1985). However it has large equivalent width narrow lines and its
absolute magnitude is below the limit for quasars used by V\'eron--Cetty
\& V\'eron (1996). It has a broad component at H$\alpha$, possibly
double peaked (Eracleous \& Halpern 1994), present also at H$\beta$. We
therefore list it as a broad line radio galaxy.

Table~\ref{obs} lists the observed objects and the parameters for the
observations, which were performed with EFOSC1 on the ESO 3.6m
telescope in La Silla in March 1995 for all objects except for 3C 327
and 0806-10 (also named 3C 195, but not a 3CR object). 
3C 327 was observed with EFOSC1 in June 1994. 
0806-10 was observed with ISIS on the William
Herschel Telescope on La Palma (Tinbergen \& Rutten 1992) in March 1994.
Details of the spectropolarimetric mode of EFOSC1
and of the observing procedure and data reduction are given by di Serego
Alighieri \& Walsh (1995). The
spectrograph slit width was 2 arcsec in all cases and was aligned with
the optical elongation of the object, when this is clearly asymmetric. EFOSC1 and
ISIS use a beamsplitting polarization analyzer to separate the ortogonal
polarization states and a half--wave plate to rotate the polarization
direction. For each object we obtained one or two sets of 4 spectral
frames taken
with the half--wave plate in position angle $0^{\circ}$, $22.5^{\circ}$,
$45^{\circ}$ and $67.5^{\circ}$. Each frame has two spectra of the
object with perpendicular polarization and several sky spectra. After
bias subtraction and flat--fielding, the object and sky spectra were
extracted from the frames and wavelength calibrated, followed by sky
subtraction and flux calibration, using observations of
spectrophotometric standard stars (Stone \& Baldwin 1983, Oke 1990). 
The portion of the slit used for object extraction was selected to
include all line emission and its length is listed in Table~\ref{obs}. The
polarization was analyzed with a set of procedures developed in
collaboration with J. Walsh (see also Walsh 1992). In order to achieve a
significant S/N ratio, the object spectra were summed over wavelength
bins selected to separate the main emission lines and line--free continuum
sections, and to avoid sky lines. We note that it is crucial to perform
this rebinning {\it before} the computation of the U and Q Stokes
parameters, in order to avoid a bias error due to the non normal error
distribution of the Stokes parameters (di Serego Alighieri 1997).
Statistical errors were derived by propagating the poissonian photon
noise in the object and sky spectra along the procedure used to obtain
the polarization parameters. The zero point offset in the polarization
position angle was obtained by observations of polarization standard
stars (Turnshek et al. 1990, Schmidt et al. 1992).

\begin{table*}
\caption{The results of spectropolarimetry}
\label{res}
\[
\begin{tabular}{llllllll}
\hline
\noalign{\smallskip}
Object & $P_{[OIII]}$ & $\theta_{[OIII]}$ & $P_{cont.}$ & $\theta_{cont.}$ & $f_{[OIII]}$ & E.W. & $E^a_{B-V}$\\
& (\%) & (deg.) & (\%) & (deg.) & ($10^{-16}erg s^{-1} cm^{-2}$) & (\AA) &\\
\noalign{\smallskip}
\hline
\noalign{\smallskip}
3C 227 & 1.27$\pm$0.52 & 15$\pm$7 & 1.90$\pm$0.12 & 50$\pm$3 & 586 & 96 & 0.006\\
3C 327 & $<1.0$ & & 0.63$\pm$0.33 & 3$\pm$15 & 891 & 208 & 0.105 \\
0806$-$10 & 1.65$\pm$0.31 & 147$\pm$4 & 3.92$\pm$0.35 & 130$\pm$3 & 1207 & 362 & 0.099 \\
1549$-$79 & 3.54$\pm$0.70 & 50$\pm$4 & 2.92$\pm$0.27 & 49$\pm$3 & 252 & 238 & 0.128 \\
3C 273$^b$ & $<2.3$ & & 0.23$\pm$0.03 & 82$\pm$3 & 2828 & 12 & 0.001 \\
3C 196.1 & $<7.2$ & & 0.83$\pm$0.21 & 130$\pm$10 & 38 & 23 & 0.055 \\
3C 180 & 1.11$\pm$0.49 & 109$\pm$8 & 0.59$\pm$0.37 & 113$\pm$15 & 396 & 317 & \\
1151$-$34 & $<3.2$ & & $<1.4$ & & 147 & 110 & 0.091 \\
1355$-$41 & $<3.6$ & & 1.22$\pm$0.17 & 108$\pm$3 & 267 & 18 & 0.063 \\
\noalign{\smallskip}
\hline
\end{tabular}
\]
\begin{list}{}{}
\item[$^{\rm a}$] Galactic extinction from Burstein \& Heiles (1982)
\item[$^{\rm b}$] For 3C 273 all [OIII] measurements refer to the 5007\AA~line only
\end{list}

\end{table*}

\section{Results}

\begin{figure*}
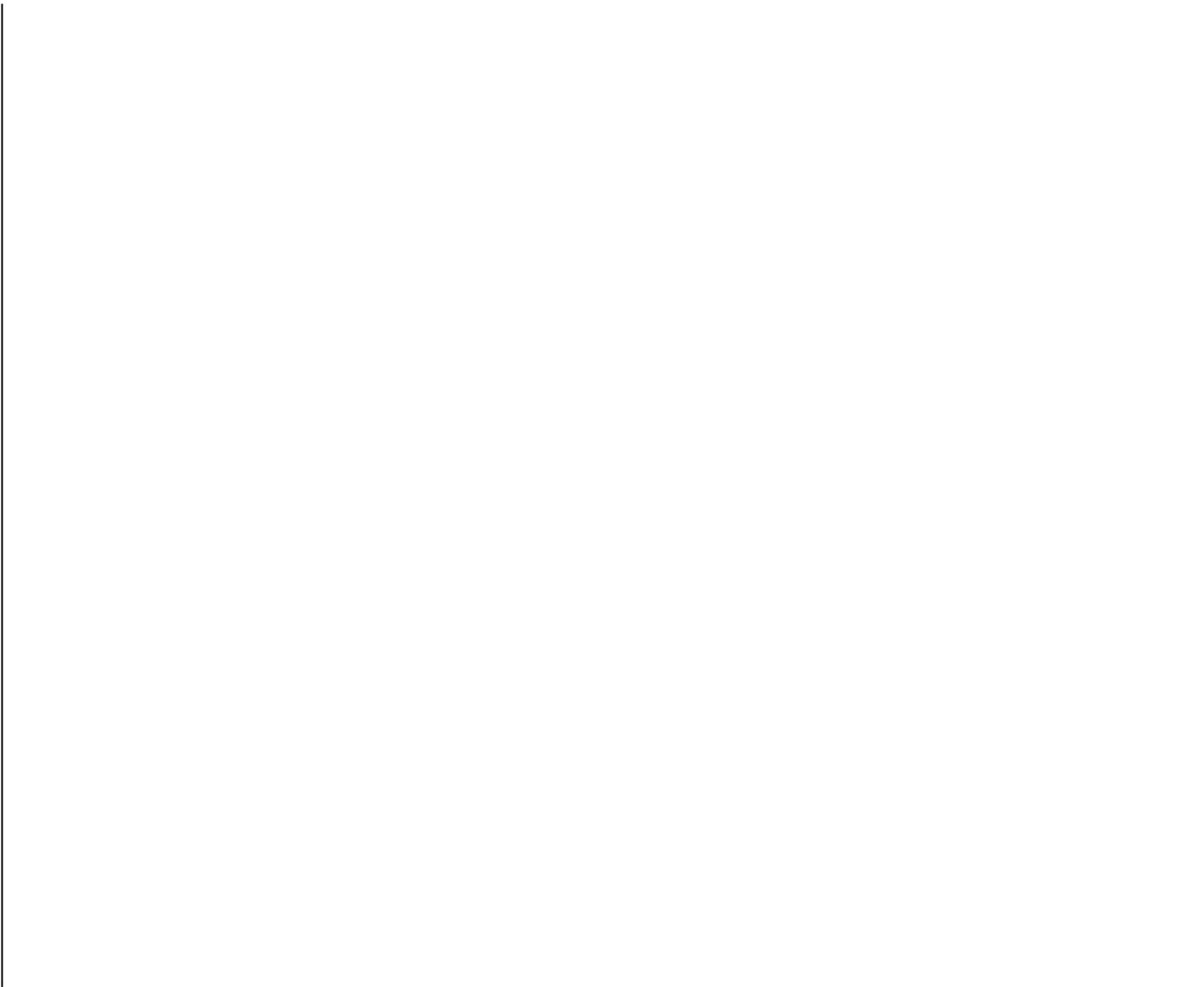

\rule{0.4pt}{15cm}
\caption{
The total flux spectra (top) and the polarized flux spectra (bottom) 
of the observed AGN. The
latter ones have been rebinned to 15\AA ~ to decrease noise. Spurious
features due to cosmic rays and to imperfect sky line subtraction are
marked with asterisks.
}
\label{fig2}
\end{figure*}

Of the results of our spectropolarimetry we present and discuss here
only those that are relevant for the anisotropy in the [OIII] emission.
The polarization of the continuum will be discussed in a separate
paper. An effective way to detect polarized line emission is to look for
the line in the polarized flux spectrum. This method provides a visual
impression of the line polarization and of its reliability, although it does
not yield the most usefull measurement. Figure~\ref{fig2} gives both the total flux
spectra and the polarized flux spectra for our sample of radio sources.
The [OIII]4959,5007 lines are clearly detected in the polarized spectra
of the RG
0806--10 and 1549--79; they are seen with some significance in the RG 3C
327, 3C 180 and in the BLRG 3C 227; they are not present in the RQ 3C
273, 1355--41 and in the BLRG 1151--34. We remark that there is no sign
of [OII]3727 in the polarized flux spectra, except possibly in
1549--79.

In order to give a quantitative measure of the line polarization we have
obtained the emission line fluxes for [OIII]4959,5007
and for [OII]3727 from the object spectra extracted from the
original 2--dimensional spectra, using a simple continuum subtraction
and line integration. Then the polarization parameters are computed from
the individual line fluxes. We have performed the continuum subtraction
in an automatic way for all the individual spectra of each object after
the selection of the continuum sampling regions, to avoid as
much as possible the effects of a subjective continuum subtraction on
the polarization. The errors were estimated taking into account all sources of
poissonian noise, including the line flux, and the subtracted continuum and 
sky. These
were propagated using a Monte Carlo simulation, similar to that described
by Fosbury et al. (1993). Table~\ref{res} lists the line polarization which we
have obtained for the sum of the two [OIII] lines at 4959 and 5007
\AA, together with the polarization of the continuum around 5000 \AA~ in
the rest frame. The degree of polarization has been corrected for the
bias expected for a definite positive quantity at low S/N (e.g. Wardle
\& Kronberg 1974). We also give the flux and the equivalent width for the
two [OIII] lines.

The results of the line polarization analysis are consistent with the
visual impression gathered from the plots of the polarized flux in
Fig.~\ref{fig2}: [OIII] polarization is measured with good accuracy
(P/$\sigma_P > 5$) for the RG 0806--10 and 1549--79, with marginal
accuracy (P/$\sigma_P \sim 2.5$) for the RG 3C 227 and 3C 180, while
P/$\sigma_P$ is smaller than 2 for the other objects and we can only
give upper limits to the polarization, corresponding to a 95.5\% degree
of confidence. No measurable polarization of [OII]3727 is
detected, although given the low S/N in the blue part of the spectrum
this result is not of high significance. It is interesting to notice
that the continuum around 5000 \AA~is polarized at a significant level 
in all objects except 1151--34.

Regarding the orientation of the polarization we first remark that there is a
significant difference in the orientation of the [OIII] and continuum 
polarization in 3C 227
and in 0806--10, while the two polarizations are parallel in the other 2
RG with clearly detected [OIII] polarization, namely 1549--79 and 3C~180.
The position angle in the optical listed in Table~\ref{obs} refers to the major
axis of the
innermost structures in broad band images (de Koff et al. 1996, Cimatti
\& di Serego Alighieri 1995), with the exception of 3C~227
for which we list the orientation of the bright [OIII] bar, since the
broad band continuum structure is round (Prieto et
al. 1993). The direction of the [OIII] polarization is perpendicular to
the [OIII] bar in 3C 227, to the elongation of the innermost isophote of
the H$\alpha$ image by Cimatti \& di Serego Alighieri (1995) in
0806--10 (i.e. 3C~195), and to the inner opposite--cones structure in
3C~180 (de Koff et al. 1996). On the other hand the direction of the
[OIII] polarization does not bear any special relation with the outer
elliptical isophotes, nor with the inner radio jet (Murphy et al. 1993) in
1549--79. However for this RG we do not have an emission line image nor
a high resolution image showing the inner structure. 
The continuum polarization around 5000\AA~is perpendicular to the
optical broad band major axis for all RG, except for the BLRG 3C~227 and for 
1549--79.

\section{Discussion}

Since the detected line polarization is of the order of a few percent
only, we cannot exclude {\it a priori} that it could be due to
trasmission through aligned dust grains. We have first examined the
influence of interstellar polarization in our Galaxy and exclude that it
is the dominant contributor to the detected line polarization for the
following reasons: a) galactic interstellar polarization would affect in
the same way the line and the continuum while we observe a different
degree and orientation in 3C 227 and in 0806--10; b) the detected line
and continuum polarization are perpendicular to well defined structures
in the environment of the AGN itself, while such a systematic trend is not expected for
galactic polarization; c) the maximum galactic interstellar
polarization expected from the galactic extinction, according to the
empirical relation ($P_{max}(\%)\le 9E_{B-V}$), is much smaller than the
observed polarization in 3C 227 and 1549--79 (Table~\ref{res} lists the galactic
extinction $E_{B-V}$ from Burstein \& Heiles 1982); d) the observed
polarization rises in the blue for all RG and is therefore inconsistent
with the Serkowski curve peaking around 5500\AA, expected for
interstellar polarization (Serkowski et al. 1975).

It is more difficult to rule out a contribution from the interstellar
polarization in the AGN host galaxy, since in this case we cannot
exclude that the lines of sight to the continuum and to the line
emission go through a different local ISM and are polarized differently,
and that the dust grains are aligned with local structures. Indeed the
polarization of the broad line RG 3C~109 at $z=0.307$, where [OIII] is
polarized at a much lower level than the continuum and broad lines, has
been interpreted as due to transmission through aligned dust grains
within the host galaxy (Goodrich \& Cohen 1992). However in this case
the polarization of the continuum follows well a Serkowski curve with a maximum around
5000\AA~ in the rest frame, and its direction is not related to the
structure of the source: it is at 60$^{\circ}$ from
the inner optical axis of the HST image (de Koff et al. 1996) and at
42$^{\circ}$ from the VLBI axis. For our sample, the perpendicularity of
the polarization with structures in the host galaxy and the
inconsistency of the observed polarization with a Serkowski curve make
it very unlikely that transmission through local dust is the main
polarizing mechanism.

The only other possible mechanism for producing net [OIII] line
polarization also in integrated light
is anisotropic scattering, which occurs when a substantial
fraction of the [OIII] emission is hidden from direct view but can
escape in other directions along which it is scattered toward us, or
when the scattering material is distributed anisotropically around most
of the [OIII] emitting region, even if this is not obscured. The
anisotropic obscuration and scattering
mechanism has been used to explain the presence of a polarized type 1
spectrum in Seyfert 2 and radio galaxies in the framework of the AGN
unification (Antonucci, 1993). In particular in distant powerful radio galaxies
the polarization has been found to be perpendicular to the elongated
optical structure, which is aligned with the radio axis (e.g. Cimatti et
al. 1993). The idea is that powerful RG harbour a quasar, whose strong continuum
and broad lines are hidden by obscuration. The obscuring material is
distributed (e.g. in a torus) such that the quasar radiation can escape
only along two opposite directions (or cones) aligned with the radio axis,
and then it is scattered in our direction with a net
polarization even in integrated light. If
the structure of the obscuring material and of the [OIII] emitting
region are such that the latter is at least partially hidden from direct
view, then [OIII] would be polarized in a direction approximately perpendicular
to the optical and radio axis. The exact correspondence of the direction
of polarization of [OIII] and of the continuum is not required, since the
two emitting regions can have a different location behind the obscuring
material and therefore a slightly different mean scattering direction.
The [OIII] polarization results on our sample are all consistent with
anisotropic scattering. Of the two possible mechanisms for producing the
anisotropy in the scattering we favour anisotropic obscuration over an
anisotropic distribution of the scattering material for similarity with
the mechanism producing the anisotropic scattering of the continuum and
broad lines, which must be obscured. However, contrary to what happens
to the nuclear continuum and broad lines, which are only visible through
scattering in many RG, a fraction of the [OIII] emission in our RG must
be seen directly, since the line polarization is generally lower than
that of the continuum, particularly if the latter is corrected for
dilution by stellar light.

In order to evaluate the fraction $f_{[OIII]_s}$ of the observed [OIII] emission which 
is scattered, we assume that the intrinsic polarization of the
scattered [OIII] is equal to that of the scattered
continuum. This assumption is reasonable if the scattering material is
the same for the line and for the continuum and if the scattering
geometries are not too different. Then
$$ f_{[OIII]_s} = f_{cont._s} {{P_{[OIII]}}\over{P_{cont.}}} $$
where $P_{[OIII]}$ and $P_{cont.}$ are the observed polarization for the
[OIII] lines and for the surrounding continuum respectively, and
$f_{cont._s}$ is the fraction of the continum which is scattered. The
latter can be estimated from the stellar fraction $f_*$, in the
assumption that the only dilution of the polarization at 5000\AA~is from stars
($f_{cont._s} = 1-f_*$). We have estimated $f_{[OIII]_s}$ only for
0806-10 and for 1549-79, where line and continuum polarization are
measured with a good accuracy. The stellar fraction at 5000\AA,
estimated from a fit of the stellar features (i.e. the CaII H and K
lines and the G band), with a synthetic stellar
spectrum are 0.55 and 0.85 respectively for the 2 RG. Then
$f_{[OIII]_s} = 0.22, 0.18$ for 0806-10, 1549-79. Although the
parameters are uncertain and there are several simplifying assumptions, if a
fifth of the observed [OIII] emission is scattered and if the scattering
efficiency is 2-3\% (see Fig. 7 of Manzini \& di Serego Alighieri 1996),
then the total [OIII] emission, if viewed directly, would be about 6 to 10
times larger than the observed one, consistent with the ratio of
[OIII] luminosity of quasars and radio galaxies as given by JB90 
and by our analysis of the southern 2Jy RG. Therefore our results bring
the [OIII] test of JB90 back in the realm of the schemes unifying
quasars and radio galaxies.

They also help in refining the unification picture. If a large fraction
of [OIII] around a radio loud AGN is emitted anisotropically and the
anisotropy is caused by obscuration, then the conditions for [OIII]
emission must exist also in regions within the obscuring material (e.g.
inside
the torus). In particular the density must be lower than the [OIII]5007
critical density ($7.9\times 10^5$ cm$^{-3}$). Although data with a higher
UV sensitivity than ours are necessary to exclude [OII]3727
polarization, the results of HBF93 suggest
that this line is emitted isotropically.
Then regions with a density below the [OII]3727 critical density
($3.0\times 10^3$ cm$^{-3}$) are not found within the obscuring material.

If the [OIII]5007 line is radiated by material at or below the critical
density for collisional de-excitation, the observed luminosity allows a
lower limit to be placed on the size of the emitting region. Where $O^{2+}$
is the dominant ionization state and the electron temperature is close
to $10^4$ K, the radius of a spherical emitting region in units of 10pc is
given by

$$r_{10} \sim 0.9 L_{42}^{1/3} f_{-5}^{-1/3} n_c^{-2/3} \zeta^{-1/3}$$

where $L_{42}$ is the line luminosity in units of $10^{42}$ erg s$^{-1}$,
$f_{-5}$ is the volume filling factor in units of $10^{-5}$ (van Breugel
1988),
$n_c$ is the electron density in units of the [OIII] critical density and
$\zeta$ is the oxygen abundance in units of the solar value.

Although such a calculation presupposes some knowledge of the filling
factor of the gas, the indicated sizes are typically somewhat smaller
than the small-scale nuclear disks seen in HST images of some nearby
radio galaxies, eg. 3C 270 with $r < 100pc$ (Jaffee et al. 1993). The
[OII] line however, because of its much lower critical density, must be
radiated by gas occupying a considerably larger volume only an
insignificant fraction of which would be obscured by such nuclear
structures.

The idea of a partial line obscuration for [OIII] but not for [OII] is
also consistent with the results of Baker (1997) on a complete
sample of quasars: she finds that the negative correlation between
emission line equivalent width and radio core dominance is much stronger
for [OIII] than for [OII].

\section{Conclusions}

We have looked for anisotropies of the [OIII] emission in radio loud AGN
by examining the [OIII] line polarization in a sample of 7 powerful radio
galaxies and 2 quasars in the redshift range $0.07\leq z\leq 0.35$. We
find that the [OIII]5007,4959 lines are significantly polarized in 4 of
the radio galaxies. In at least 3 of them the direction of [OIII]
polarization is perpendicular to the axis of the extended line
emission. Although some contribution from polarization by interstellar
dust in the host galaxy cannot be excluded, our results strongly suggest
that a fraction of the [OIII] line emission is not seen directly in radio
galaxies, but is scattered toward us after anisotropic obscuration by
the same material which hides the quasars postulated by the Unified
Model. For the 2 radio galaxies where line polarization is observed with
the highest accuracy, we estimate that the [OIII] line radiation which
is emitted behind the obscuring material can be about 6 to 10 times larger
that the directly observed one, thereby explaining within the framework
of the Unified Model the higher [OIII] luminosity of quasars with
respect to radio galaxies. Spectropolarimetric data with higher
sensitivity and spectral resolution are necessary first to check whether
anisotropic scattering is less important for [OII], as
expected from the equivalent [OII] luminosity of quasars and radio
galaxies. Secondly, important information can be obtained on the kinematics of the line
emitting gas and of the scattering material by a comparison of the line
profiles in the total flux spectra and in the polarized flux
spectra, in analogy to what has been done for planetary nebulae (e.g.
Walsh \& Clegg 1994). Finally, differences in the geometric
distribution of the obscured regions emitting the continuum, the broad lines and
the narrow [OIII] lines can be examined in more detail. Such differences
are seen in planetary nebulae and in the AGN Cygnus A (Ogle et al.
1997).

\begin{acknowledgements}
We thank the staff of the ESO/La Silla Observatory and of the ING on La
Palma for support during the observations, Jeremy Walsh for illuminating
discussions on polarimetry, and Ski Antonucci, the referee, for useful
comments. One of us (R.H.) is grateful to the
Osservatorio Astrofisico di Arcetri for hospitality during this work.
This research has made use of the Simbad
database, operated at CDS, Strasbourg.
\end{acknowledgements}


\begin{thebibliography}{}


\bibitem[]{}
Antonucci R. R. J., 1984, ApJ, 278, 499

\bibitem[]{}
Antonucci R., 1993, ARA\&A, 31, 473

\bibitem[]{}
Baker J.C., 1997, MNRAS, 286, 23

\bibitem[]{}
Burstein D., Heiles C., 1982, AJ, 87, 1167

\bibitem[]{}
Cimatti A., di Serego Alighieri S., 1995, MNRAS, 273, L7

\bibitem[]{}
Cimatti A., di Serego Alighieri S., Fosbury R. A. E., Salvati M., 
Taylor D., 1993, MNRAS, 264, 421

\bibitem[]{}
de Koff S., Baum S. A., Sparks W. B., Biretta J., Golombek D.,
Macchetto F., McCarthy P., Miley G. K., 1996, ApJS, 107, 621

\bibitem[]{}
di Serego Alighieri S., Cimatti A., Fosbury R. A. E., 1994, ApJ,
431, 123

\bibitem[]{}
di Serego Alighieri S., Walsh J. R., 1995, Spectropolarimetry with
EFOSC1. In: Benvenuti P. (ed.) Calibrating and understanding HST and
ESO instruments'', ESO Conf.
and Workshop Proc. No. 53, Garching bei M\"unchen, p. 71 

\bibitem[]{}
di Serego Alighieri S., 1997, Polarimetry with large telescopes. In:
Rodriguez de Espinosa J. M. et al. (eds.) Instrumentation for large
telescopes, Cambridge University Press, Cambridge (in press)

\bibitem[]{}
Eracleous M., Halpern J. P., 1994, ApJS, 90, 1

\bibitem[]{}
Fosbury R. A. E., Cimatti A., di Serego Alighieri S. 1993, The
Messenger, 74, 11

\bibitem[]{}
Goodrich R. W., Cohen M. H., 1992, ApJ, 391, 623

\bibitem[]{}
Hes R., Barthel P. H., Fosbury R. A. E., 1993, Nature, 326, 362

\bibitem[]{}
Jackson N., Browne I. W. A., 1990, Nature, 343, 43

\bibitem[]{}
Jaffee W., Ford H. C., Ferrarese L., van den Bosch F., O'Connel
R. W., 1993, Nature, 364, 213

\bibitem[]{}
Keel W.C., de Grijp M.H.K., Miley G.K., Zheng W., 1994, A\& A, 283, 791

\bibitem[]{}
Lawrence A., 1987, PASP, 99, 309

\bibitem[]{}
Lawrence A., 1991, MNRAS, 252, 586

\bibitem[]{}
Manzini A., di Serego Alighieri S., 1996, A\& A, 311, 79

\bibitem[]{}
Murphy D.W. and the SHEVE Team, 1993, 2.3 GHz VLBI images of southern
hemisphere radio galaxies and quasars. In: Davis R. J. \& Booth R. S.
(eds.) Cambridge University Press, Cambridge, p. 243

\bibitem[]{}
Ogle P. M., Cohen M. H., Miller J. S., Tran H. D., Fosbury R. A.
E., Goodrich R. W., 1997, ApJ (in press)

\bibitem[]{}
Oke J. B., 1990, AJ, 99, 1621

\bibitem[]{}
Prieto M. A., Walsh J. R., Fosbury R. A. E., di Serego Alighieri S.,
1993, MNRAS, 263, 10

\bibitem[]{}
Schmidt G. D., Elston R., Lupie O. L., 1992, AJ, 104, 1563

\bibitem[]{}
Serkowski K., Mathewson D. S., Ford V. L., 1975, ApJ, 196, 261

\bibitem[]{}
Spinrad H., Djorgovski S., Marr J., Aguilar L., 1985, PASP, 97, 932

\bibitem[]{}
Stone R. P. S., Baldwin J. A., 1983, MNRAS, 204, 347

\bibitem[]{}
Tadhunter C. N., Morganti R., di Serego Alighieri S., Fosbury R. A. E.,
Danziger I. J., 1993, MNRAS, 263, 999

\bibitem[]{}
Tinbergen J., Rutten R., 1992, A User Guide to WHT
Spectropolarimetry, User Manual No. XXI, Isaac Newton Group, La Palma

\bibitem[]{}
Turnshek D. A., Bohlin R. C., Williamson II R. L., Lupie O. L.,
Koornneef J., Morgan D. H., 1990, AJ, 99, 1243

\bibitem[]{}
van Breugel W. J. M., 1988, Extended optical line emission in radio
galaxies. In: Meisenheimer K. \& R\"oser H.--J. (eds.)
Hot spots in extragalactic radio sources, Springer, Berlin, p. 121

\bibitem[]{}
V\'eron--Cetty M.P., V\'eron P., 1996, A catalogue of quasars
and active nuclei (7th Edition), ESO Scientific Report No. 17, Garching
bei M\"unchen

\bibitem[]{}
Wall J. V., Peacock J. A., 1985, MNRAS, 216, 173

\bibitem[]{}
Walsh J. R., 1992, The reduction of spectropolarimetry data. In: Grosb\o l 
P. J. \& de Ruijsscher R. C. E. (eds.) 4$^{th}$ ESO/ST--ECF Data Analysis
Workshop, ESO Conf. and Workshop Proc. No. 41, Garching bei M\"unchen, p. 53

\bibitem[]{}
Walsh, J. R., \& Clegg, R., 1994, MNRAS, 268, L41

\bibitem[]{}
Wardle J. F. C., Kronberg P. P., 1974, ApJ, 194, 249 

\end{thebibliography}
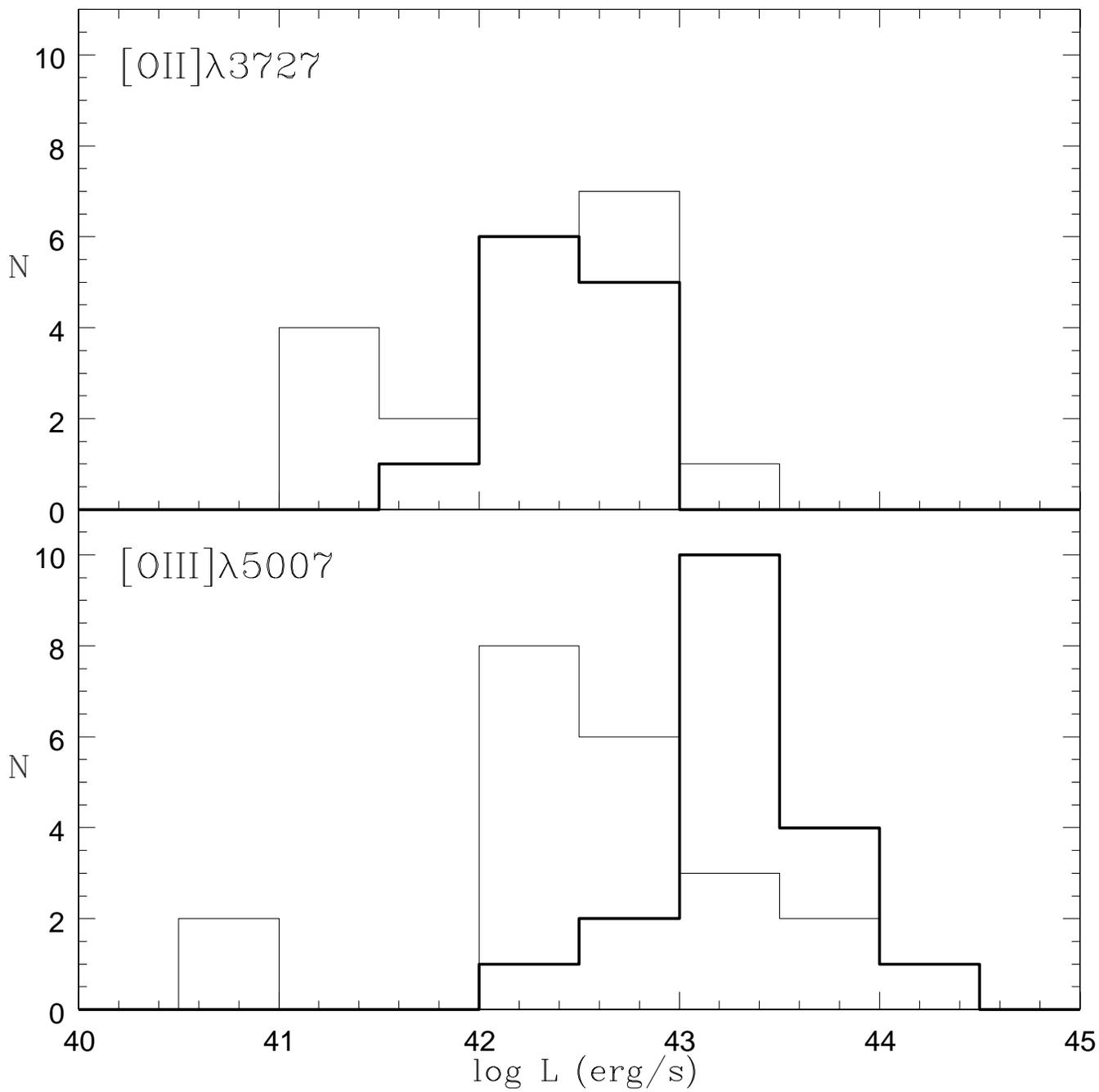
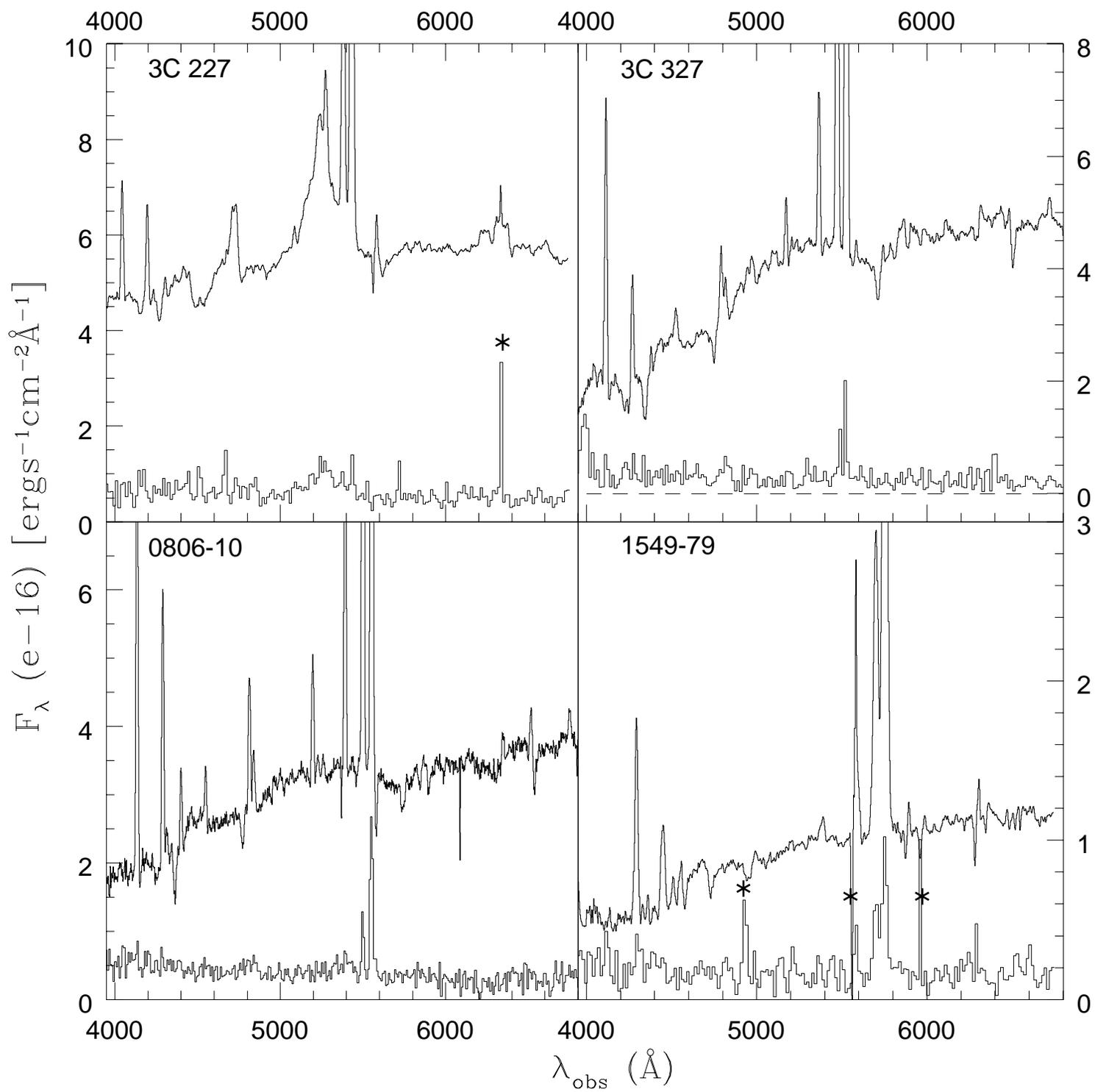
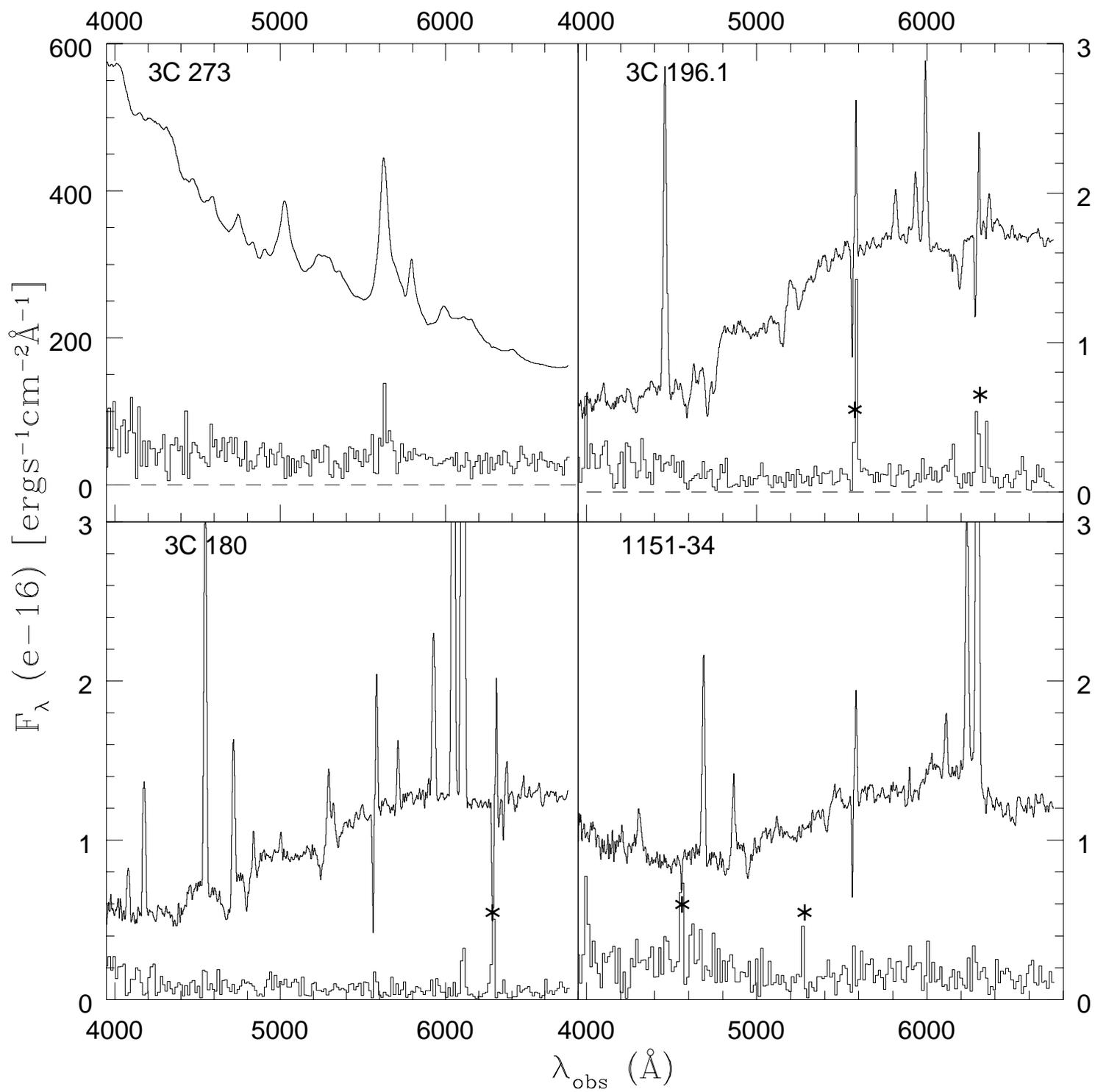
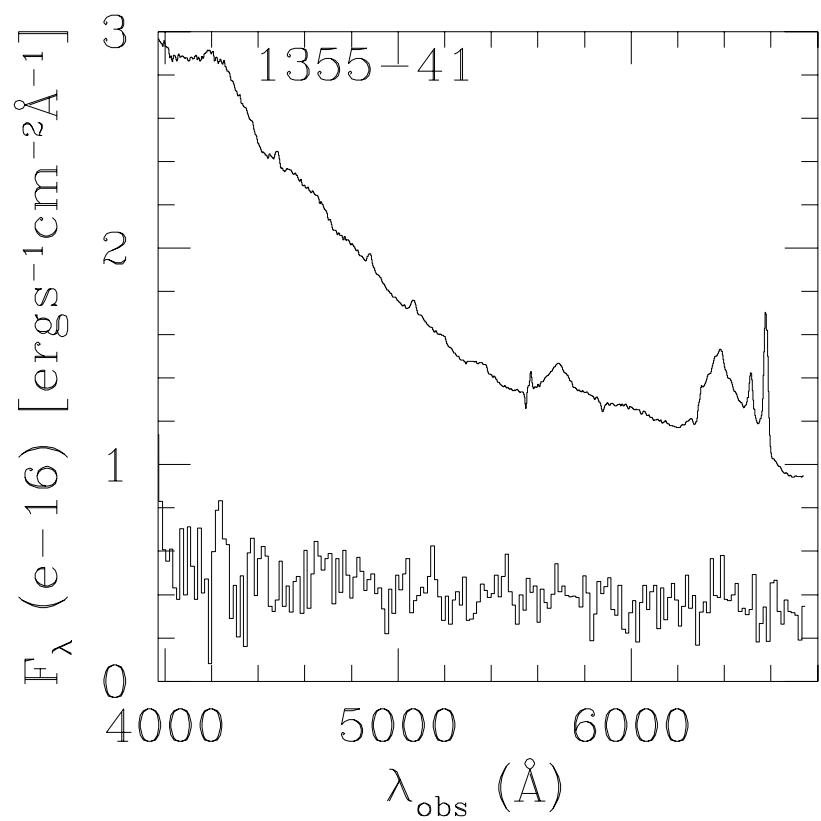
\end{document}